\documentclass[superscriptaddress]{revtex4-1}
\usepackage[pdftex]{graphicx}
\usepackage{float}
\usepackage{amsmath}
\usepackage{bm}

\begin{document}

\title{Active matter clusters at interfaces.}

\author{Katherine Copenhagen}
\affiliation{University of California Merced, Merced CA}
\author{Ajay Gopinathan}
\affiliation{University of California Merced, Merced CA}

\begin{abstract}
Collective and directed motility or swarming is an emergent phenomenon displayed by many self-organized assemblies of active biological matter such as clusters of embryonic cells during tissue development, cancerous cells during tumor formation and metastasis, colonies of bacteria in a biofilm, or even flocks of birds and schools of fish at the macro-scale. Such clusters typically encounter very heterogeneous environments. What happens when a cluster encounters an interface between two different environments has implications for its function and fate. Here we study this problem by using a mathematical model of a cluster that treats it as a single cohesive unit that moves in two dimensions by exerting a force/torque per unit area whose magnitude depends on the nature of the local environment.  We find that low speed (overdamped) clusters encountering an interface with a moderate difference in properties can lead to refraction or even total internal reflection of the cluster. For large speeds (underdamped), where inertia dominates, the clusters show more complex behaviors crossing the interface multiple times and deviating from the predictable refraction and reflection for the low velocity clusters. We then present an extreme limit of the model in the absense of rotational damping where clusters can become stuck spiraling along the interface or move in large circular trajectories after leaving the interface.  Our results show a wide range of behaviors that occur when collectively moving active biological matter moves across interfaces and these insights can be used to control motion by patterning environments.
\end{abstract}

\maketitle

\section*{Introduction.}

Swarming is a widespread biological phenomenon characterized by long range order emerging in a system from local interactions between agents\cite{Sumpter2006}, such as a swarm of flies\cite{Okubo1974, Kelley2013}, or flock of birds\cite{Bialek}.  Typically, a group of individual organisms self-organize to form a cohesive cluster with directed motility in a spontaneously chosen consensus direction, for example a school of fish\cite{Herbert-read2011}, cluster of cells during tumor growth, tissue development and repair\cite{Vicsek,Szabo2006}, or herd of wildebeests\cite{Gueron1993}.  These types of swarming systems will often encounter a change in the environment, such as a flock of birds flying into a cloud or area of lower air temperature, or a cluster of tumor cells invading different tissue types\cite{Friedl2012}.  Single cells have been shown to change their speed and direction when crossing sharp interfaces \cite{Dokukina2010}.  Swarms can also use collective dynamics to find and localize themselves to preferred niches or microenvironments. For example it has been observed that golden shiners, \textit{Notemigonus crysoleucas}, which prefer low lighting, will spend more time in dark areas if it is part of a school due to cooperative sensing capabilities of the group\cite{Berdahl2013}.  Bacteria have also been shown to take less time to reach a target in the presence of noisy chemical concentration gradients in the environment when they are part of a cluster\cite{Shklarsh2011}.  Also \textit{e. coli.} clusters modify their own environment by secreting chemicals to create an environmental change between the inside and outside of the cluster in order to trap the \textit{e. coli.} and maintain clustering behaviors\cite{Mittal2003}.  It is therefore important to understand the effects of spatial environmental changes on a cohesive swarming group.  In this paper we investigate finite swarming clusters moving through heterogeneous environments where agents change their speeds by exerting different forces within each environment. Such changes could arise from the agent's sensing and response to a variety of environmental factors such as temperature, substrate stiffness, or chemical composition.  So, how is the path that a swarm takes affected by the presence of a boundary between different environments, and how does that depend on the properties of the cluster and the environmental change?

Agent based models\cite{Vicsek1995} and hydrodynamic continuum models\cite{Toner1998} have been used with great success to model natural collective systems\cite{Sumpter2006, Couzin2003}, and reveal phases and transitions which emerge from the active, far from equilibrium nature of these systems\cite{Couzin2002,Guillaume2004}.  Agent based models are implemented by defining a set of interactions and update rules for individuals and then letting the system evolve in time, such systems show phase transitions driven by a wide variety of quantities including noise, density\cite{Zafeiris}, environmental disorder\cite{Chepizhko2013b}, behavioral heterogeneities\cite{McCandlish2012,Baglietto2013}, and cohesive interaction details \cite{DOrsogna2006,Gazi2004}.  They have also been used to study how swarms use cooperation to achieve specific goals which are useful to biological systems, such as cooperative decision making\cite{Couzin2005}, agent segregation\cite{Belmonte2008}, and obstacle avoidance\cite{Quint}.  Hydrodynamic continuum models on the other hand do not treat each agent as an individual, but instead study an average alignment and density profile within the system\cite{Toner2005}. Hydrodynamic swarming models have established active matter as a type of nonequilibrium complex fluid\cite{Toner1995,Marchetti2013} and provided a unified framework to study phase transitions\cite{Levine2000,Toner2005}, instabilities\cite{Bertin2009} and pattern formation\cite{Liu2013} in active systems.

To answer the question of swarms crossing from one environment into another, we utilize a simplified model for a swarm that assumes a polarized ordered state with velocities correlated across the system\cite{Cavagna2010} and finite system size, which allows us to examine the overall behaviors of the swarm crossing an interface without necessitating details about individual agent motion as in agent based models, or infinite system sizes as in hydrodynamic models. We consider a swarm as a single cohesive disk-shaped unit, or cluster, each individual agent within the swarm manifests as a force per unit area applied to the disk in the direction of polarization.  We then allow the cluster to cross an interface between two differing environments where the portion of the swarm in each environment may apply stronger or weaker forces to the cluster depending on the substrate, resulting in a torque and therefore a curved trajectory.  We can then map the trajectories of the cluster and measure the resulting relationships between the cluster motion before and after crossing the interface. We find four catagories for cluster behavior which depend on two important cluster parameters. The two important parameters are cluster speed, i.e. low speed (overdamped) vs. high speed (underdamped) clusters, and the ratio of the rotational damping to translational friction. It is to be noted that though a prescribed amount of rotational damping arises from the translational friction on the disk, the ratio is a parameter that could be smaller (or bigger) depending on whether the cluster exerts torques to maintain (or resist) rotation.

Our model has predictive capabilities for determining the curved path of a cell cluster at an interface between substrates.  The results of which could be extended to suggest possible methods of patterning substrates to direct cell cluster motion.  Other regimes of our model provide insight into the behaviors of faster swarms, such as bird flocks and fish schools, moving between heterogeneous environments.  Finally, for swarms that exert torques to maintain turning, we show that swarms can display circular paths and trapping at interfaces.  

\section*{The Model.}

\begin{figure}
\centering
\includegraphics[width=0.75\textwidth]{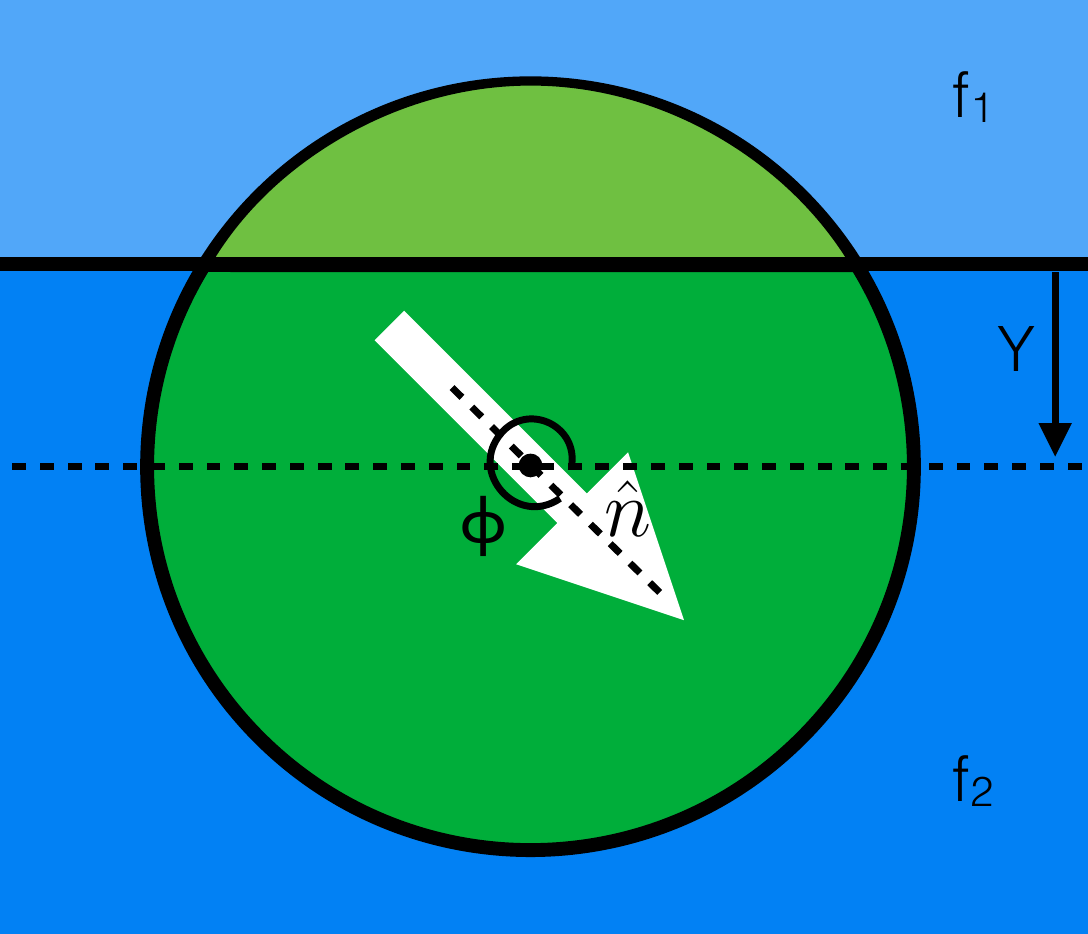}
\caption{View from above showing the two regions one above (light green) and below (dark green) the interface between two different substrates shown in light and dark blue respectively.  Each region of the cluster propels itself along the polarization direction ($\hat{n}$ at an angle of $\Phi$ with the positive $x$-axis) with a force per unit area of $f_1$ for the top region and $f_2$ for the bottom region.  The cluster center is at a height $Y$ above the interface (in the diagram shown $Y$ is negative because the center of the cluster is below the interface).}
\label{diag}
\end{figure}

We use a mathematical model for a swarming cluster (Fig.~\ref{diag}) that treats it as a single cohesive unit that moves on a two dimensional substrate by exerting a force per unit area in some cluster polarization direction (at an angle $\Phi$ with respect to the positive $x$-axis).  We then examine a single cluster moving across an interface between two different substrates where the area of the cluster contained on substrate $1$ (shown in light green above the solid horizontal line in Fig.~\ref{diag}) exerts a force per unit area of $f_1$ and the portion on substrate $2$ (shown by the dark green area in Fig.~\ref{diag}) exerts a force per unit area of $f_2$.  The force applied on the cluster by every portion of the cluster is in the direction of polarization with a magnitude that depends on the areas within each substrate and the substrate dependent forces. The cluster also experiences a friction-like damping force resisting translational motion.  Utilizing these details we can calculate a force on the cluster at any height $Y$ above a horizontal substrate interface (See appendix for details).

If the cluster is polarized at an angle which is not normal to the interface there is an asymmetry of the forces on either side of the interface, which results in a torque on the direction of polarization of the cluster which in turn rotates the direction of the force on the cluster and can result in a curved trajectory or a bend in the path of the cluster as it crosses the interface.  This torque can be calculated from the force per unit area of the cluster on each substrate along with the distance from the cluster center.  From the derived expressions for force and torque (see appendix) we can find the non-dimensionalized equations of motion for the cluster shown in eqs.\ref{eomphi}-\ref{eomy}, where $v_1$ and $v_2$ are the equilibrium speeds of the cluster in the top and bottom substrates respectively, and $C$ is the ratio of the angular damping to the translational friction on the disk.  In the non-dimensionalized forms of the equations shown below the translational/frictional damping constant is incorporated into the equilibrium speeds $v_1$ and $v_2$.

\begin{equation}
\frac{d^2{\Phi}}{dT^2}=\frac{4}{3\pi}\sqrt{1-Y^2}(1+Y^2)(v_2-v_1)\cos(\Phi)-C\frac{d\Phi}{dT}
\label{eomphi}
\end{equation}
\begin{equation}
\frac{d^2X}{dT^2}=\Big(1/2(v_1+v_2)+1/\pi(v_1-v_2)\big(\arcsin(Y)+Y\sqrt{1-Y^2}\big)\Big)\cos(\Phi)-\frac{dX}{dT}
\label{eomx}
\end{equation}
\begin{equation}
\frac{d^2Y}{dT^2}=\Big(1/2(v_1+v_2)+1/\pi(v_1-v_2)\big(\arcsin(Y)+Y\sqrt{1-Y^2}\big)\Big)\sin(\Phi)-\frac{dY}{dT}
\label{eomy}
\end{equation}

We can then use finite difference methods to solve these equations of motion and examine the system subjected to different substrates and initial conditions.  In the model the unit of length is set by the cluster radius, and the unit of time is set by the time taken for the cluster to accelerate from rest to the fraction $(1-1/\mathrm{e})\sim0.63$ of its equilibrium speed.  The active nature of the system allows the angular damping and translational friction to not be equal, implying $C\neq1$.  Physically we can expect $C>1$ for situations when swarms exert torques to resist turning, and $C<1$ for cases where swarms exert torques to maintain or persist in turning as seen in systems like the persistent turning walker model for \textit{K. Mugil} fish\cite{Gautrais2009}.  How, then, does the incident angle and ratio of equilibrium speeds in each substrate affect the transmitted angle of the swarm? What is the effect of exerted torques which promote turning?

\section*{Results.}
\begin{figure}
\centering
\includegraphics[width=0.75\textwidth]{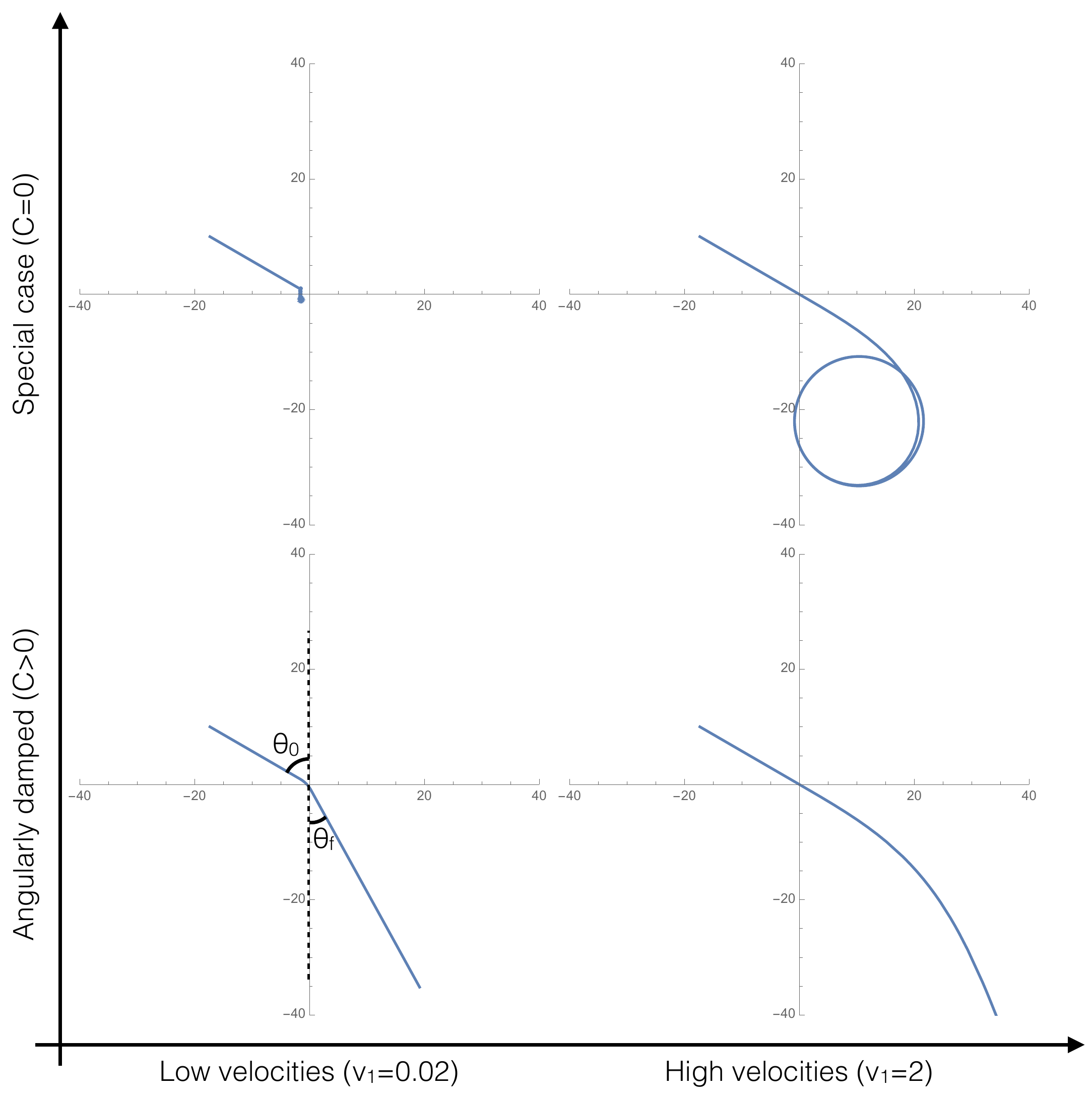}
\caption{Representative trajectories of the system in each of the four limitting behaviors of the system. In all four cases $v_1=2v_2$.}
\label{phasespace}
\end{figure}

 Fig.~\ref{phasespace} shows the four characteristic trajectories for limits of the two important parameters: low velocities (overdamped, friction dominated behaviors), high velocities (underdamped, inertia dominated behaviors), and clusters with angular damping ($C>0$).  At the end we consider a special limiting case where $C=0$, relevant/applicable to agents that apply torques to promote turning which exactly cancel out any angular damping, resulting in unique cluster behaviors.
 
\subsection{Cluster trajectories.}

\begin{figure}
\centering
\includegraphics[width=0.75\textwidth]{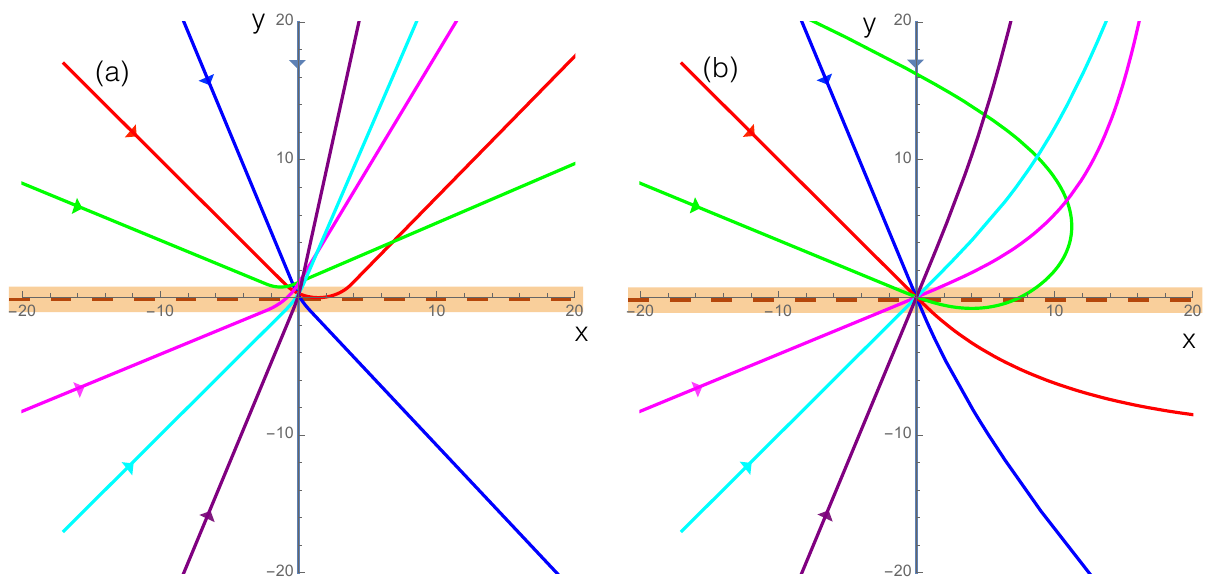}
\caption{Example trajectories of the cluster, each color is a different incident angle where the incoming cluster is shown by the straight lines entering from the left of each plot ($x < 0$) directed towards the origin, and the clusters are in contact with the interface, which is at $y=0$ (maroon dashed line), when the trajectory is within the orange band. The substrate on the bottom half ($y<0$) has twice the equilibrium speed as the top substrate ($v_2=2v_1$) in both cases. (a) For low speeds, where friction dominates cluster motion, the cluster curves while it is in contact with the interface (orange band), and once it leaves the interface it has a well defined straight path along some rotated polarization direction. (b) High velocities, when inertia dominates, result in the cluster starting to curve when it comes into contact with the interface and then large sweeping curves away from the interface as the momentum of the cluster causes it to persist along the previous direction before gradually adjusting to the new substrate.}
\label{trajs}
\end{figure}

The overall behaviors of the cluster in the presence of angular damping fall into two different categories: refraction/reflection (Fig~\ref{trajs}(a)), and large sweeping curves (Fig~\ref{trajs}(b)). In both cases the direction that the cluster actively propels itself in ($\hat{n}$ at angle $\Phi$) can only accelerate due to torques experienced while it is in contact with the interface.  This means that while the cluster is on a single substrate the angular speed of $\hat{n}$ can only decrease due to rotational damping.

In the low velocity cases ($v_1=0.01$ in Fig.~\ref{trajs}(a)) friction dominates the cluster motion and the cluster moves in a direction parallel to the active propulsion of the cluster ($\vec{v}||\hat{n}$) at nearly all times. In this friction dominated limit, the system can refract or reflect off of the interface and only comes into contact with the interface once allowing us to measure the incident and refracted angle and make predictions about the behaviors and interactions of the cluster with an interface.

The high velocity case ($v_1=5$ in Fig.~\ref{trajs}(b)) is characterized by the fact that the cluster's translational inertia dominates the direction of cluster motion ($\vec{v}$), for an extended period of time after crossing the interface.  In this high velocity case, as the cluster leaves the interface it will move in a direction which is not necessarily parallel to the direction of $\hat{n}$, resulting in the apparent angular acceleration away from the interface. However, after some time the system will reach equilibrium where $\vec{v}||\hat{n}$ and the cluster will move in a straight line in some well defined direction.  In this case the momentum of the cluster carries it away from the interface even if the direction of $\hat{n}$ is such that the cluster should be propelled towards the interface meaning that the cluster can rotate and return to the interface a finite number of times before reaching a stable straight trajectory within a single substrate.  This high velocity (inertia dominated) case could model the motion of a bird flock crossing an interface between hotter and colder air where their momentum will continue to carry the flock in one direction and the speed and trajectory will gradually stabilize as the flock adjusts to the new environment, potentially traveling in a different direction from that with which it entered.  

\subsection{Refraction and reflection at an interface.}

\begin{figure}[H]
\centering
\includegraphics[width=0.75\textwidth]{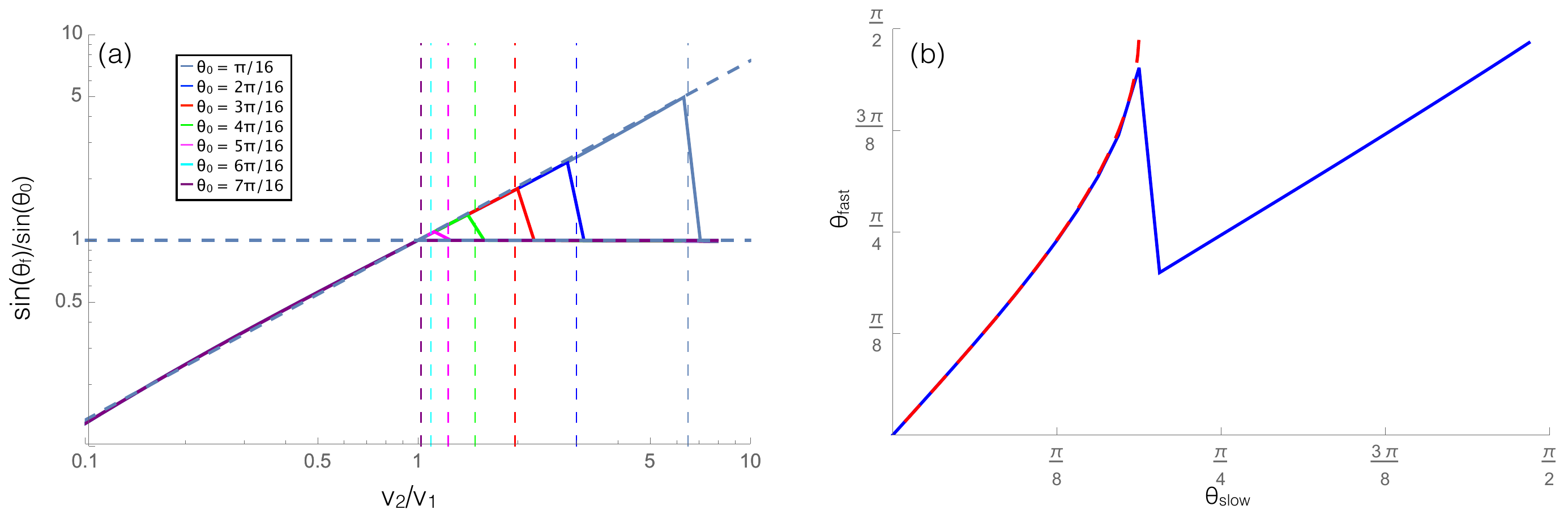}
\caption{(a) The ratios of the sines of the incident ($\theta_0$) and refracted ($\theta_f$) angles plotted against the ratios of the equilibrium speeds in each material, with $v_1=0.001$.  The vertical lines show the value of velocity ratio where the refracted angle would become greater than $\pi/2$, resulting in a reflection, for each initial angle shown in the legend.  When the cluster is reflected (right side of the vertical dashed lines) the horizontal line at $\sin\theta_f/\sin\theta_0=1$ shows that the reflected angle is equal to the incident angle. (b) The angle that the trajectory makes (with respect to the normal to the interface) on the faster ($\theta_{fast}$) and slower ($\theta_{slow}$) substrates for a cluster moving from the slower substrate to the faster one shown in blue, and the similar angles for a cluster moving from the faster to the slower substrate shown in red dashed line.  In both these cases $v_1 = 0.001$, and $v_2=0.002$, and $C=1$. }
\label{snell}
\end{figure}

In this section we examine the case where the velocity on each substrate is low enough that friction dominates, and the cluster experiences rotational damping ($C>0$).  The cluster in this regime is representative of slow moving natural swarming systems, including cell clusters, and bacteria swarms. A cluster approaching an interface at an angle will begin to turn when it comes into contact with the interface and when contact with the interface is lost it will travel in a straight line in a new direction.  We can then measure the resulting angle and the incident angle (labeled $\theta_f$ and $\theta_0$ respectively in Fig.~\ref{phasespace}) and compare them to the velocity ratio on the two substrates.  To do this we plot the ratio of the sines of the refracted and incident angles against the ratio of the equilibrium velocities on each substrate.  Fig.~\ref{snell} (a) shows the relationship between these ratios for many initial angles (see legend), along with a fit line to the numerical data for the refraction angle which leads to the predicted reflections. Applying a similar fit to systems with different values of $C$ leads to the relationship shown in Eq.~\ref{snellslaw}. This relation bears a remarkable resemblance to Snell's law for optical paths except with an exponent of $0.87/C$ instead of unity.  Additionally, when moving from a slower substrate to a faster one the cluster will be reflected off the interface if the transmitted angle should be $\pi/2$ or greater, i.e. total internal reflection.  The velocity ratio where reflection should begin to occur is shown as vertical dashed lines for different incident angles according to the color legend.  The predicted reflections align well with the plotted data, where the ratio of the sines for each incident angle begins to reflect at the same point as the vertical dashed lines.  The reflection is shown by the horizontal dashed line at $1$, which implies that the reflected angle is equal to the incident angle beyond the velocity ratio which would result in $\theta_f >\pi/2$.  

\begin{equation}
\frac{\sin{\theta_2}}{\sin{\theta_1}} = \Big(\frac{v_2}{v_1}\Big)^{0.87/C}
\label{snellslaw}
\end{equation}

The trajectories of these clusters are also reversible in time as can be seen in Fig.~\ref{snell}(b) which shows the cluster trajectory angle with respect to normal on the faster substrate plotted against the angle of the trajectory on the slower substrate.  The blue solid line shows the case where the cluster is moving from the slower substrate onto the faster one, and drops down to the diagonal line when reflection begins to occur.  The red dashed line shows the angles for the faster to slower case.  The overlap of these two curves shows that whether moving from faster to slower or vice versa the angles depend only on the substrates and angles and not on the initial substrate of the cluster.  The refraction of the cluster into the slower medium while approaching at a large angle from the faster medium is similar to the reported behavior of the golden shiners that collectively turn into darker regions where they move slower\cite{Berdahl2013}.

\subsection{Fast swarms and swarms with exerted torques to promote turning.}

\begin{figure}
\centering
\includegraphics[width=0.75\textwidth]{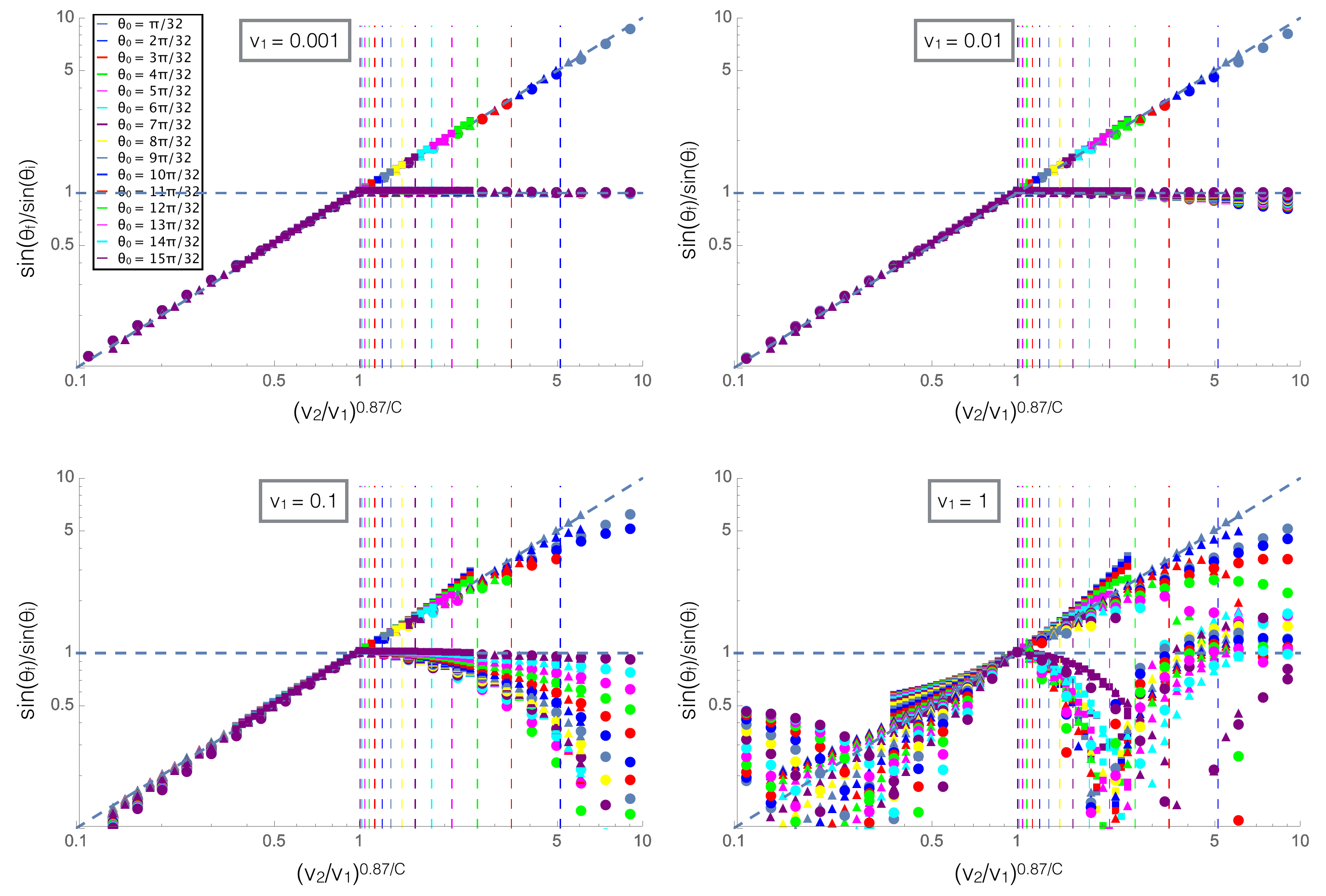}
\caption{The ratios of the sines of the initial and final trajectory angles measured with respect to the normal to the interface.  The velocity on the top substrate ($v_1$) is labeled for each plot and the color legend for incident angle is consistent across all four plots.  Squares are for $C=2$, triangles for $C=1$ and circles for $C=0.5$.  The diagonal line is a fit for the low velocity case and the vertical lines are predicted values for the reflections to begin for each incident angle shown by color according to the legend.}
\label{fastswarms}
\end{figure}

In the case where the velocity on each substrate is high enough for interia to dominate cluster motion, the cluster will still adjust its direction as it passes through the interface but its own momentum will carry it straight across the interface before slowly adjusting it's angle depending on the rotations caused by the interface which we can see qualitatively in Fig.~\ref{trajs}(b).  This results in the cluster often returning to the interface and interacting with it multiple times making the refracted angle vary from the predictions made for the friction dominated case.  

Fig.~\ref{fastswarms} shows the ratio of the incident and refracted angles of clusters versus the ratio of velocities scaled by angular damping similar to Fig.~\ref{snell}(a).  In this plot the collapse at low velocities for different values of $C$ is shown by the shapes of the markers, $C=0.5$ for the circles, $C=1$ (rotational damping arising from translational friction only) for the triangles and $C=2$ for the squares, the shapes and color legend are consistent across all four plots shown. We can see that at high velocities the reflected angle varies greatly from the low velocity case where reflected angle was equal to incident angle.  This is due to the fact that the cluster passes through the interface before turning around and then passes through the interface again resulting in multiple interactions with the interface and final reflected angles that don't relate in a well defined way to the incident angles.  The refracted angle also differs from the low velocity case due to the effects of momentum and inertia at high velocities, the trajectories become complicated and cease to follow a well defined relationship.  This behavior could be related to the complex fluid-like motions of starling flocks as they swoop and change directions in large arcing trajectories, possibly due to changing environments from flying over trees or through different altitudes and air temperatures.

\begin{figure}
\centering
\includegraphics[width=0.75\textwidth]{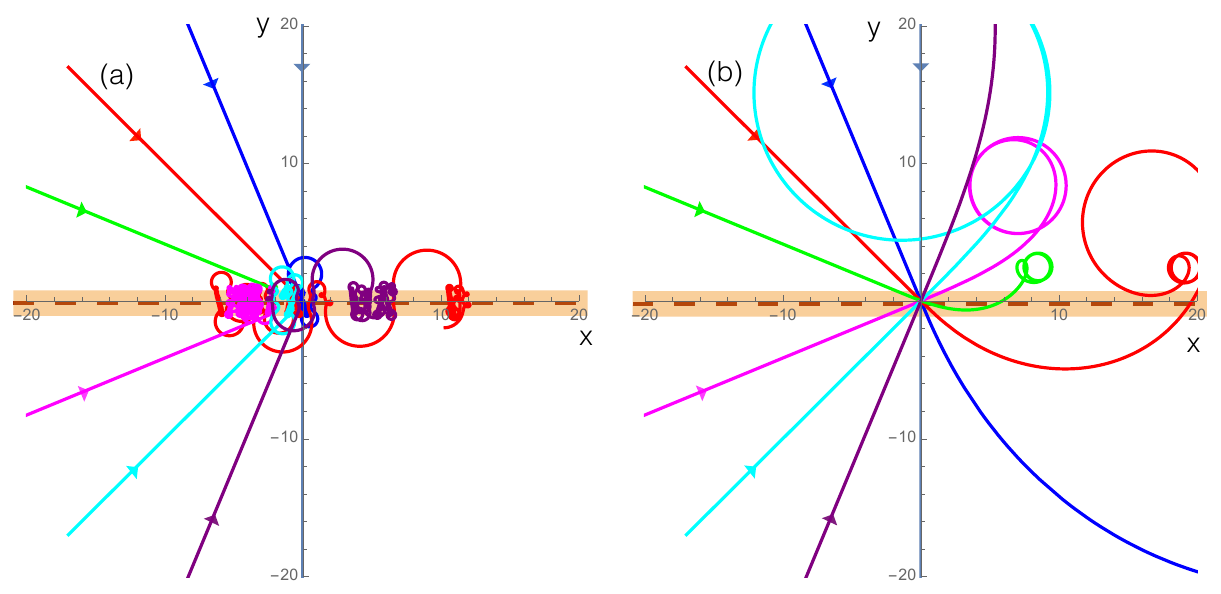}
\caption{(a) Trajectories for the $C=0$ case with low velocities.  $v_1=0.001$ and $v_2=0.002$, the incident trajectories are the straight lines coming from the left. (b) Same as (a) but for the high velocity case where $v_1=5$ and $v_2=10$. The cluster is in contact with the interface while the trajectory is shown in the orange band.}
\label{trajs2}
\end{figure}

Finally we consider a special case of a swarm that actively promotes rotations by exerting a torque that exactly cancels out any rotational resistance from friction-like damping (resulting in $C=0$).  In this case two very different behaviors emerge.  The first is in the low velocity case where friction dominates the translational motion, though angular momentum is always important due to the fact that $C=0$ independent of velocity.  In this case the cluster will again begin to rotate as it crosses the interface, and experience no angular acceleration once it loses contact with the interface, however it will continue to rotate along a circular curve and return to the interface some time later.  This causes the cluster to become trapped at the interface by always rotating around and returning to the interface without being able to escape as seen in Fig.~\ref{trajs2}(a).  The second case is for high velocities where the inertia of the cluster will carry it past the interface and it may return to the interface or escape but the initial crossing of the interface starts the cluster rotating and the cluster will move in a circle somewhere away from the interface typically on the slower substrate as seen in Fig.~\ref{trajs2}(b).

\section*{Discussion.}

Collective directed motility is a phenomenon that is widespread in biological systems including cell clusters during tissue development and tumor formation, as well as bacterial biofilms and flocks of birds. In these types of systems it is reasonable to assume that clusters of collectively moving agents move through changing environments, be it a change in air temperature, stiffness in substrate or any other change which could result in speed change.  We have used a model which treats a swarming cluster as a single cohesive unit with a preferred direction, to examine the effects of a swarm moving across an interface between two environments due to a change in speed that occurs in each separate environment.  

We found that clusters can display different broad behaviors.  The most applicable of which is for slow moving swarms with some angular damping.  In this case a swarming cluster approaching an interface at an angle will undergo some form of refraction or reflection resulting in a new direction which is predictable by a simple relationship between incident and refracted angle and the ratio of the equilibrium speeds on each substrate.  Clusters in this regime can also display total internal reflection at the predictable angle where the refraction angle would exceed $\pi/2$.  This regime of our model could represent cell clusters on changing substrates and our predictions could be used to pattern a substrate to direct cluster motion along a desired path.

When the velocities of the cluster is increased on both substrates the trajectories gradually diverge from the predicted refraction angles and reflections found for low velocities where friction dominates cluster motion. This is due to the inertia of the clusters carrying it quickly across the interface with the inability of the cluster to change directions at a comparable rate.  As the cluster velocities continue to increase the trajectories become sensitive to initial conditions and rotational damping, due to the cluster interacting multiple times with the interface or spending less time in contact with the interface than necessary for the cluster to adjust its direction according to the torques present.  These kinds of clusters display broad sweeping curved trajectories away from the interface and could provide insight into the impressive collective motion seen in starling flocks and fast moving fish schools, such as sardines.  

In the special case where the cluster experiences no angular damping, possibly due to swarms exerting a torque which counter-acts the friction-like resistance that is present in the translational motion, the cluster will either become stuck on the interface for low velocities or stabilize in circular trajectories on a single substrate for high velocities.  These types of behaviors may be desirable for certain systems, and our predictions could be used to engineer cell clusters, or robot algorithms to follow desired paths.  Trapping a cell cluster on an interface could be useful for separating cells or subjecting them to specific conditions which could then be applied at the interface more easily to cell clusters as individual cells will not display the same kinds of collective modes at an interface.  Cell cluster motion has been shown to be controllable by hard boundaries\cite{Doxzen2013}, and our results suggest that a similar strategy could be used with softer substrate interfaces which can be crossed to manipulate cell cluster behaviors.

Our results show possible predictive capabilities for slow moving clusters such as cells or bacteria moving across changing substrates, as well as possible insight into ongoing questions such as the behaviors of starling flocks and fish schools as they spiral and curve while they span and cross interfaces between changing environments.  Additionally our results suggest possible mechanisms for directing collective systems by way of changing environmental conditions.

\begin{acknowledgments}

This work was partially supported by National Science Foundation (NSF) grant EF-1038697 (to A.G.), a James S. McDonnell Foundation Award (to A.G.) and an NSF-IGERT graduate fellowship (to K.C.)

\end{acknowledgments}
]
\newpage
\newpage
 \section*{Appendix.}

\begin{figure}
\centering
\includegraphics[width=0.75\textwidth]{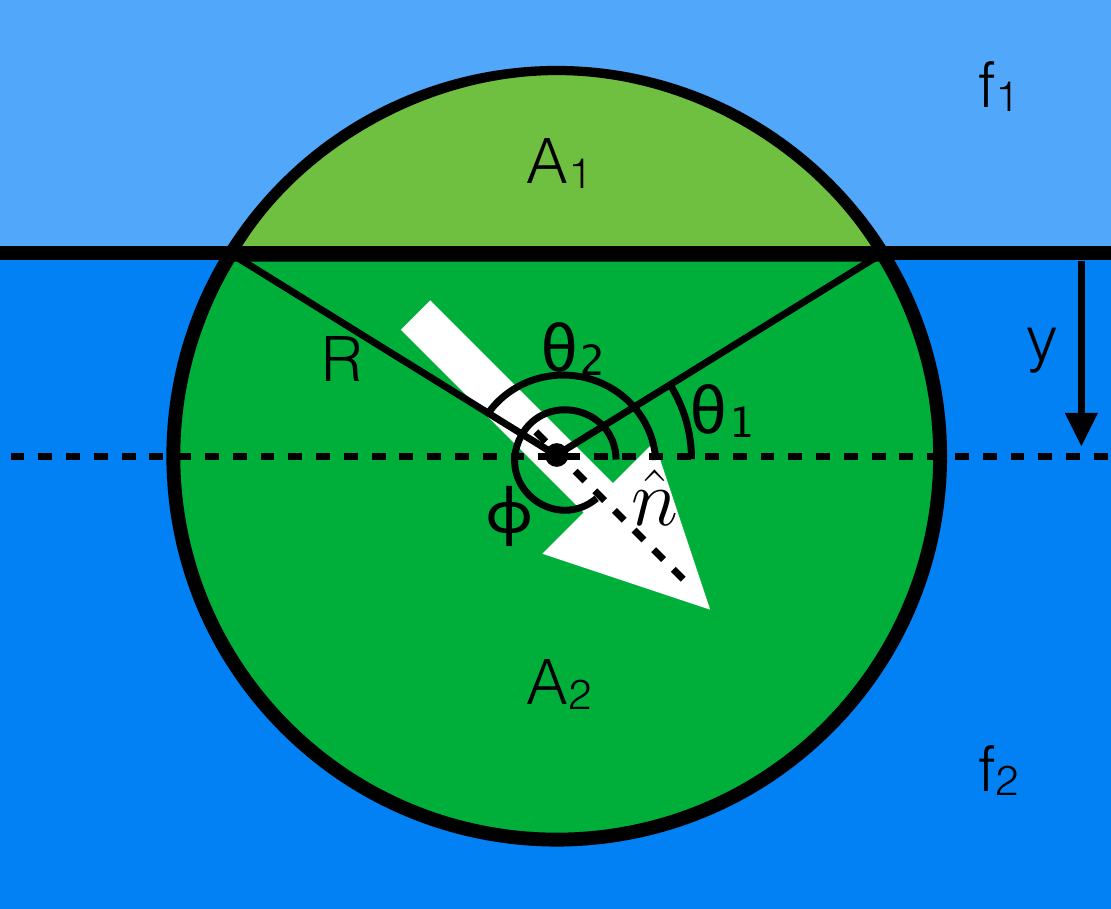}
\caption{Diagram of cluster setup, the parameters are described in the appendix text.}
\label{diag2}
\end{figure}

We use a mathematical model for the cluster that treats it as a single cohesive unit that moves on a two dimensional substrate by exerting a force and torque per unit area whose magnitude depends on the nature of the substrate.  Here we examine the effects of a single cluster moving across an interface between two different substrates where each portion of the cluster exerts a force per unit area depending on which substrate it is on.  The force exerted on the cluster in direction of polarization ($\hat{n}$) is equal to the substrate dependent force per unit area ($f_{1(\text{or }2)}$) multiplied by the area of the cluster on each respective substrate ($A_{1(\text{or }2)}$), with damping constant $b$.

\begin{equation}
{\bf F} = (f_1A_1+f_2A_2)\hat{n} - b\vec{v}
\end{equation}

The area of the partial circles within each region at any heigh $y$ above the substrate interface can be calculated as shown below, where $R$ is the radius of the cluster, and $\theta_1$ and $\theta_2$ are the angles to the intersections of the interface line and the rim of the cluster as shown in Fig.~\ref{diag2}.  Integrating and simplifying results in Eq.~\ref{Force} as the net force on the cluster with radius $R$ at height $y$ above the interface.  

\begin{displaymath}
{\bf F} = \Bigg(f_1 \int_{\theta_1}^{\theta_2}  \int_{\frac{y}{\sin{\theta}}}^R  r dr  d\theta + f_2 \int_{\theta_2}^{2\pi+\theta_1}\int_{\frac{y}{\sin{\theta}}}^R rdrd\theta\Bigg)\hat{n} - b\vec{v}
\end{displaymath}
\begin{displaymath}
{\bf F} = \Bigg(f_1 \int_{\arcsin{(-\frac{y}{R}})}^{\pi-\arcsin{(-\frac{y}{R}})}\frac{1}{2}\Big(R^2-\frac{y^2}{\sin^2\theta}\Big) d\theta + f_2 \int_{\pi-\arcsin{(-\frac{y}{R}})}^{2\pi+\arcsin{(-\frac{y}{R}})} \frac{1}{2}\Big(R^2-\frac{y^2}{\sin^2\theta}\Big) d\theta \Bigg)\hat{n} - b\vec{v}
\end{displaymath}
\begin{displaymath}
{\bf F} = \frac{1}{2}\Bigg(f_1\Big(R^2\theta+y^2\cot{\theta}\Big)_{\arcsin{(-\frac{y}{R})}}^{\pi-\arcsin{(-\frac{y}{R})}}+f_2\Big(R^2\theta+y^2\cot{\theta}\Big)_{\pi-\arcsin{(-\frac{y}{R}})}^{2\pi+\arcsin{(-\frac{y}{R}})}\Bigg)\hat{n} - b\vec{v}
\end{displaymath}
\begin{multline*}
{\bf F} = \frac{1}{2}\Bigg(f_1\Big(R^2(\pi-\arcsin(-\frac{y}{R}))+y^2\cot(\pi-\arcsin(-\frac{y}{R}))-(R^2(\arcsin(-\frac{y}{R}))+y^2\cot(\arcsin(-\frac{y}{R})))\Big) \\ +f_2 \Big(R^2(2\pi+\arcsin(-\frac{y}{R}))+y^2\cot(2\pi+\arcsin(-\frac{y}{R}))-(R^2(\pi-\arcsin(-\frac{y}{R}))+y^2\cot(\pi-\arcsin(-\frac{y}{R})))\Big)\Bigg)\hat{n} - b\vec{v}
\end{multline*}
\begin{displaymath}
{\bf F} = \frac{1}{2}\Bigg(f_1\Big(R^2(\pi+2\arcsin(\frac{y}{R}))+2y\sqrt{R^2-y^2}\Big)+f_2\Big(R^2(\pi-2\arcsin(\frac{y}{R}))-2y{\sqrt{R^2-y^2}})\Big)\Bigg)\hat{n} - b\vec{v}
\end{displaymath}
\begin{displaymath}
{\bf F}=\frac{1}{2}R^2\Bigg(f_1\Big(\pi+2\arcsin(\frac{y}{R})+2\frac{y}{R}\sqrt{1-(y/R)^2}\Big)+f_2\Big(\pi-2\arcsin(\frac{y}{R})-2\frac{y}{R}\sqrt{1-(y/R)^2}\Big)\Bigg)\hat{n}-b\vec{v}
\end{displaymath}

The expression for force on the cluster due to the active force per unit area exerted by the cluster and damping is shown below.

\begin{equation}
{\bf F}=\frac{1}{2}R^2\Big(\pi(f_1+f_2)+2(f_1-f_2)\big(\arcsin(\frac{y}{R})+\frac{y}{R}\sqrt{1-(y/R)^2}\big)\Big)\hat{n}-b\vec{v}
\label{Force}
\end{equation}

In addition to the substrate dependent force applied on the cluster, the polarization direction of the cluster will also be subject to a torque due to the asymmetry of the variable forces on each area of the cluster.  The net torque on the cluster can be calculated by integrating the torque per unit area over the portions of the cluster on each substrate similar to the calculation of force above.  The torque per unit area is due to the force per unit area at each point on the disk, i.e. $f_{1(\text{or }2)}\hat{n}\times\vec{r}$, with angular damping $-c{\boldsymbol \omega}$.

\begin{equation}
{\boldsymbol \tau}=\int_{A_1}f_1\hat{n}\times\vec{r}dA+\int_{A_2}f_2\hat{n}\times\vec{r}dA-c{\boldsymbol \omega}
\end{equation}

The area of the portions of the cluster on each substrate is integrated as in the force case above, and simplified to get the expression for torque on the cluster shown in Eq.~\ref{Torque}.

\begin{displaymath}
{\boldsymbol \tau}=\int_{\theta_1}^{\theta_2}\int_{\frac{y}{\sin\theta}}^R f_1\hat{n}\times\vec{r} rdr d\theta+\int_{\theta_2}^{2\pi+\theta_1}\int_{\frac{y}{\sin\theta}}^R f_2\hat{n}\times\vec{r} rdr d\theta-c{\boldsymbol \omega}
\end{displaymath}
\begin{displaymath}
{\boldsymbol \tau}=\Big(\int_{\theta_1}^{\theta_2}\int_{\frac{y}{\sin\theta}}^R f_1r\sin(\theta-\phi)rdrd\theta+\int_{\theta_2}^{2\pi+\theta_1}\int_{\frac{y}{\sin\theta}}^R f_2r\sin(\theta-\phi)rdrd\theta\Big)\hat{z}-c{\boldsymbol \omega}
\end{displaymath}
\begin{displaymath}
{\boldsymbol \tau}=\Big(\int_{\theta_1}^{\theta_2}f_1\sin(\theta-\phi)\frac{r^3}{3}\Big|_{\frac{y}{\sin\theta}}^Rd\theta+\int_{\theta_2}^{2\pi+\theta_1}f_2\sin(\theta-\phi)\frac{r^3}{3}\Big|_{\frac{y}{\sin\theta}}^Rd\theta\Big)\hat{z}-c{\boldsymbol \omega}
\end{displaymath}
\begin{displaymath}
{\boldsymbol \tau}=\Big(\int_{\theta_1}^{\theta_2}f_1(\sin(\phi)\cos(\theta)-\cos(\phi)\sin(\theta))(\frac{R^3}{3}-\frac{y^3}{3\sin^3\theta})d\theta+\int_{\theta_2}^{2\pi+\theta_1}f_2(\sin(\phi)\cos(\theta)-\cos(\phi)\sin(\theta))(\frac{R^3}{3}-\frac{y^3}{3\sin^3\theta})d\theta\Big)\hat{z}-c{\boldsymbol \omega}
\end{displaymath}
\begin{multline*}
{\boldsymbol \tau}=\Big(\int_{\theta_1}^{\theta_2}f_1\frac{R^3}{3}(\sin(\phi)\cos(\theta)-\cos(\phi)\sin(\theta))-y^3\frac{\sin(\phi)\cos(\theta)}{3\sin^3(\theta)}+y^3\frac{\cos(\phi)}{3\sin^2(\theta)}d\theta \\ +\int_{\theta_2}^{2\pi+\theta_1}f_2\frac{R^3}{3}(\sin(\phi)\cos(\theta)-\cos(\phi)\sin(\theta))-y^3\frac{\sin(\phi)\cos(\theta)}{3\sin^3(\theta)}+y^3\frac{\cos(\phi)}{3\sin^2(\theta)}d\theta\Big)\hat{z}-c{\boldsymbol \omega}
\end{multline*}
\begin{multline*}
{\boldsymbol \tau}=\Big(f_1\Bigg(\frac{R^3}{3}\sin(\phi)\sin(\theta)+\frac{R^3}{3}\cos(\phi)\cos(\theta)+\frac{y^3\sin(\phi)}{6\sin^2(\theta)}-\frac{y^3\cos(\phi)\cos(\theta)}{3\sin(\theta)}\Bigg)_{\arcsin(-\frac{y}{R})}^{\pi-\arcsin(-\frac{y}{R})} \\ +f_2\Bigg(\frac{R^3}{3}\sin(\phi)\sin(\theta)+\frac{R^3}{3}\cos(\phi)\cos(\theta)+\frac{y^3\sin(\phi)}{6\sin^2(\theta)}-\frac{y^3\cos(\phi)\cos(\theta)}{3\sin(\theta)}\Bigg)_{\pi-\arcsin(-\frac{y}{R})}^{2\pi+\arcsin(-\frac{y}{R})}\Big)\hat{z}-c{\boldsymbol \omega}
\end{multline*}
\begin{multline*}
{\boldsymbol \tau}=\hat{z}\Big(f_1 \times \\ \Bigg(\frac{R^3}{3}\sin(\phi)\sin(\pi-\arcsin(-\frac{y}{R}))+\frac{R^3}{3}\cos(\phi)\cos(\pi-\arcsin(-\frac{y}{R}))+\frac{y^3\sin(\phi)}{6\sin^2(\pi-\arcsin(-\frac{y}{R}))}-\frac{y^3\cos(\phi)\cos(\pi-\arcsin(-\frac{y}{R}))}{3\sin(\pi-\arcsin(-\frac{y}{R}))}\Bigg) \\-f_1\Bigg(\frac{R^3}{3}\sin(\phi)\sin(\arcsin(-\frac{y}{R}))+\frac{R^3}{3}\cos(\phi)\cos(\arcsin(-\frac{y}{R}))+\frac{y^3\sin(\phi)}{6\sin^2(\arcsin(-\frac{y}{R}))}-\frac{y^3\cos(\phi)\cos(\arcsin(-\frac{y}{R}))}{3\sin(\arcsin(-\frac{y}{R}))}\Bigg) +f_2 \times \\ \Bigg(\frac{R^3}{3}\sin(\phi)\sin(2\pi+\arcsin(-\frac{y}{R}))+\frac{R^3}{3}\cos(\phi)\cos(2\pi+\arcsin(-\frac{y}{R}))+\frac{y^3\sin(\phi)}{6\sin^2(2\pi+\arcsin(-\frac{y}{R}))}-\frac{y^3\cos(\phi)\cos(2\pi+\arcsin(-\frac{y}{R}))}{3\sin(2\pi+\arcsin(-\frac{y}{R}))}\Bigg) \\ -f_2\Bigg(\frac{R^3}{3}\sin(\phi)\sin(\pi-\arcsin(-\frac{y}{R}))+\frac{R^3}{3}\cos(\phi)\cos(\pi-\arcsin(-\frac{y}{R}))+\frac{y^3\sin(\phi)}{6\sin^2(\pi-\arcsin(-\frac{y}{R}))}-\frac{y^3\cos(\phi)\cos(\pi-\arcsin(-\frac{y}{R}))}{3\sin(\pi-\arcsin(-\frac{y}{R}))}\Bigg) \Big)\\-c{\boldsymbol \omega}
\end{multline*}
\begin{multline*}
{\boldsymbol \tau}=\hat{z}\Big(f_1 \Bigg(\frac{R^3}{3}\sin(\phi)(-y/R)+\frac{R^3}{3}\cos(\phi)(-\sqrt{R^2-y^2}/R)+\frac{y^3\sin(\phi)}{6y^2/R^2}-\frac{y^3\cos(\phi)(-\sqrt{R^2-y^2}/R)}{3(-y/R)}\Bigg) \\-f_1\Bigg(\frac{R^3}{3}\sin(\phi)(-y/R)+\frac{R^3}{3}\cos(\phi)(\sqrt{R^2-y^2}/R)+\frac{y^3\sin(\phi)}{6y^2/R^2}-\frac{y^3\cos(\phi)\sqrt{R^2-y^2}/R}{3(-y/R)}\Bigg)  \\+f_2 \Bigg(\frac{R^3}{3}\sin(\phi)(-y/R)+\frac{R^3}{3}\cos(\phi)(\sqrt{R^2-y^2}/R)+\frac{y^3\sin(\phi)}{6y^2/R^2}-\frac{y^3\cos(\phi)\sqrt{R^2-y^2}/R}{3(-y/R)}\Bigg) \\ -f_2\Bigg(\frac{R^3}{3}\sin(\phi)(-y/R)+\frac{R^3}{3}\cos(\phi)(-\sqrt{R^2-y^2}/R)+\frac{y^3\sin(\phi)}{6y^2/R^2}-\frac{y^3\cos(\phi)(-\sqrt{R^2-y^2}/R)}{3(-y/R)}\Bigg) \Big)\\-c{\boldsymbol \omega}
\end{multline*}
\begin{multline*}
{\boldsymbol \tau}=\hat{z}\Big(f_1\Bigg(-2\frac{R^2}{3}\cos(\phi)(\sqrt{R^2-y^2})-2\frac{y^2\cos(\phi)(\sqrt{R^2-y^2})}{3}\Bigg) \\+f_2 \Bigg(2\frac{R^2}{3}\cos(\phi)(\sqrt{R^2-y^2})+2\frac{y^2\cos(\phi)\sqrt{R^2-y^2}}{3}\Bigg) \Big)-c{\boldsymbol \omega}
\end{multline*}
\begin{displaymath}
{\boldsymbol \tau}=\hat{z}\frac{2}{3}\sqrt{R^2-y^2}\cos(\phi)\Big(f_1(-R^2-y^2)+f_2(R^2+y^2)\Big)-c{\boldsymbol \omega}
\end{displaymath}

The equation below shows the torque on the cluster due to the active forces on each substrate, and rotational damping.

\begin{equation}
{\boldsymbol \tau}=\hat{z}\frac{2}{3}\sqrt{R^2-y^2}\cos(\phi)(f_2-f_1)(R^2+y^2)-c{\boldsymbol \omega}
\label{Torque}
\end{equation}

From eq.~\ref{Force} for force and eq.~\ref{Torque} for torque we can write down the equations of motion for the orientation as well as the x and y coordinates of the cluster, as shown below.

\begin{displaymath}
\frac{1}{2}mR^2\ddot{\phi}=\frac{2}{3}\sqrt{R^2-y^2}\cos(\phi)(R^2+y^2)(f_2-f_1)-c\dot{\phi}
\end{displaymath}
\begin{displaymath}
m\ddot{x}=\frac{1}{2}R^2\Big(\pi(f_1+f_2)+2(f_1-f_2)\big(\arcsin(\frac{y}{R})+\frac{y}{R}\sqrt{1-(y/R)^2}\big)\Big)\cos(\phi)-b\dot{x}
\end{displaymath}
\begin{displaymath}
m\ddot{y}=\frac{1}{2}R^2\Big(\pi(f_1+f_2)+2(f_1-f_2)\big(\arcsin(\frac{y}{R})+\frac{y}{R}\sqrt{1-(y/R)^2}\big)\Big)\sin(\phi)-b\dot{y}
\end{displaymath}

We can then nondimensionalize the equations of motion through the following substitutions:

\begin{displaymath}
\phi=\Phi
\end{displaymath}
\begin{displaymath}
x=RX
\end{displaymath}
\begin{displaymath}
y=RY
\end{displaymath}
\begin{displaymath}
t=\frac{m}{b}T
\end{displaymath}

\begin{displaymath}
\frac{R^2b^2}{2m}\frac{d^2{\Phi}}{dT^2}=\frac{2}{3}R^3\sqrt{1-Y^2}(1+Y^2)(f_2-f_1)\cos(\Phi)-\frac{cb}{m}\frac{d\Phi}{dT}
\end{displaymath}
\begin{displaymath}
\frac{Rb^2}{m}\frac{d^2X}{dT^2}=\frac{1}{2}R^2\Big(\pi(f_1+f_2)+2(f_1-f_2)\big(\arcsin(Y)+Y\sqrt{1-Y^2}\big)\Big)\cos(\Phi)-\frac{Rb^2}{m}\frac{dX}{dT}
\end{displaymath}
\begin{displaymath}
\frac{Rb^2}{m}\frac{d^2Y}{dT^2}=\frac{1}{2}R^2\Big(\pi(f_1+f_2)+2(f_1-f_2)\big(\arcsin(Y)+Y\sqrt{1-Y^2}\big)\Big)\sin(\Phi)-\frac{Rb^2}{m}\frac{dY}{dT}
\end{displaymath}

\begin{displaymath}
\frac{d^2{\Phi}}{dT^2}=\frac{4mR}{3b^2}\sqrt{1-Y^2}(1+Y^2)(f_2-f_1)\cos(\Phi)-\frac{2c}{R^2b}\frac{d\Phi}{dT}
\end{displaymath}
\begin{displaymath}
\frac{d^2X}{dT^2}=\frac{mR}{2b^2}\Big(\pi(f_1+f_2)+2(f_1-f_2)\big(\arcsin(Y)+Y\sqrt{1-Y^2}\big)\Big)\cos(\Phi)-\frac{dX}{dT}
\end{displaymath}
\begin{displaymath}
\frac{d^2Y}{dT^2}=\frac{mR}{2b^2}\Big(\pi(f_1+f_2)+2(f_1-f_2)\big(\arcsin(Y)+Y\sqrt{1-Y^2}\big)\Big)\sin(\Phi)-\frac{dY}{dT}
\end{displaymath}

In the limit where the cluster is entirely on one substrate, we can compute the equilibrium speed of the cluster on each substrate by letting the acceleration go to zero, and $Y>1$ for the first substrate or $Y<-1$ for the other substrate.  Taking the real components of the equations of motion these limits of the system give us the equilibrium speeds as given below.  

\begin{displaymath}
0=\frac{mR}{2b^2}\Big(\pi(f_1+f_2)+2(f_1-f_2)(\pm\frac{\pi}{2})\Big)\hat{n}-\vec{v}
\end{displaymath}

\begin{displaymath}
v_{1/2}=\frac{\pi mR}{b^2}f_{1/2}
\end{displaymath}

Substituting equlibrium velocities into the equations of motion results in the eqs.~\ref{eomphi}-\ref{eomy}.  Additionally, the value of $c$ due only to friction-like resistance is $c=bR^2/2$.  Substituting $c=CbR^2/2$, where $C$ is the ratio of the actual rotational resistance to the frictional angular resistance, results in the following equations of motion.

\begin{equation}
\frac{d^2{\Phi}}{dT^2}=\frac{4}{3\pi}\sqrt{1-Y^2}(1+Y^2)(v_2-v_1)\cos(\Phi)-C\frac{d\Phi}{dT}
\end{equation}
\begin{equation}
\frac{d^2X}{dT^2}=\Big(1/2(v_1+v_2)+1/\pi(v_1-v_2)\big(\arcsin(Y)+Y\sqrt{1-Y^2}\big)\Big)\cos(\Phi)-\frac{dX}{dT}
\end{equation}
\begin{equation}
\frac{d^2Y}{dT^2}=\Big(1/2(v_1+v_2)+1/\pi(v_1-v_2)\big(\arcsin(Y)+Y\sqrt{1-Y^2}\big)\Big)\sin(\Phi)-\frac{dY}{dT}
\end{equation}


\begin{thebibliography}{36}

\bibitem{Sumpter2006}
D~J~T Sumpter.
\newblock {The principles of collective animal behaviour}.
\newblock {\em Philosophical Transactions of the Royal Society B}, 361:5--22,
  2005.

\bibitem{Okubo1974}
Akira Okubo and H.~C. Chiang.
\newblock {An analysis of the kinematics of swarming of Anarete pritchardi kim
  (Diptera: Cecidomyiidae)}.
\newblock {\em Researches on Population Ecology}, 16(1):1--42, 1974.

\bibitem{Kelley2013}
Douglas~H Kelley and Nicholas~T Ouellette.
\newblock {Emergent dynamics of laboratory insect swarms.}
\newblock {\em Scientific reports}, 3:1073, jan 2013.

\bibitem{Bialek}
William Bialek, Andrea Cavagna, Irene Giardina, Thierry Mora, Oliver Pohl,
  Edmondo Silvestri, Massimiliano Viale, and Aleksandra~M Walczak.
\newblock {Social interactions dominate speed control in poising natural flocks
  near criticality.}
\newblock {\em Proceedings of the National Academy of Sciences of the United
  States of America}, 111(20):7212--7, 2014.

\bibitem{Herbert-read2011}
J.~E. Herbert-Read, Andrea Perna, R.~P. Mann, T.~M. Schaerf, D.~J.~T. Sumpter,
  and A.~J.~W. Ward.
\newblock {Inferring the rules of interaction of shoaling fish}.
\newblock {\em Proceedings of the National Academy of Sciences},
  108(46):18726--18731, 2011.

\bibitem{Vicsek}
Tam{\'{a}}s Vicsek.
\newblock {Collective motion of cells : from experiments to models}.
\newblock pages 1--24.

\bibitem{Szabo2006}
B.~Szab{\'{o}}, G.J. Sz{\"{o}}ll{\"{o}}si, B.~G{\"{o}}nci, Zs. Jur{\'{a}}nyi,
  D.~Selmeczi, and Tam{\'{a}}s Vicsek.
\newblock {Phase transition in the collective migration of tissue cells:
  Experiment and model}.
\newblock {\em Physical Review E}, 74(6):1--5, 2006.

\bibitem{Gueron1993}
S~Gueron and Sa~Levin.
\newblock {Self-organization of Front Patterns in Large Wildebeest Herds},
  1993.

\bibitem{Friedl2012}
Peter Friedl, Joseph Locker, Erik Sahai, and Jeffrey~E. Segall.
\newblock {Classifying collective cancer cell invasion}.
\newblock {\em Nature Cell Biology}, 14(8):777--783, 2012.

\bibitem{Dokukina2010}
Irina~V Dokukina and Maria~E Gracheva.
\newblock {A model of fibroblast motility on substrates with different
  rigidities.}
\newblock {\em Biophysical journal}, 98(12):2794--803, jun 2010.

\bibitem{Berdahl2013}
Andrew Berdahl, Colin~J Torney, Christos~C Ioannou, Jolyon~J Faria, and Iain~D
  Couzin.
\newblock {Emergent sensing of complex environments by mobile animal groups.}
\newblock {\em Science (New York, N.Y.)}, 339(6119):574--6, 2013.

\bibitem{Shklarsh2011}
Adi Shklarsh, Gil Ariel, Elad Schneidman, and Eshel Ben-Jacob.
\newblock {Smart swarms of bacteria-inspired agents with performance adaptable
  interactions}.
\newblock {\em PLoS Computational Biology}, 7(9):e1002177, 2011.

\bibitem{Mittal2003}
Nikhil Mittal, Elena~O Budrene, Michael~P Brenner, and Alexander {Van
  Oudenaarden}.
\newblock {Motility of Escherichia coli cells in clusters formed by chemotactic
  aggregation.}
\newblock {\em Proceedings of the National Academy of Sciences of the United
  States of America}, 100(23):13259--13263, 2003.

\bibitem{Vicsek1995}
Tamas Vicsek, A~Czir{\'{o}}k, E~Ben-Jacob, I~Cohen, and O~Shochet.
\newblock {Novel type of phase transition in a system of self-driven
  particles}.
\newblock {\em Physical Review Letters}, 75(6):4--7, 1995.

\bibitem{Toner1998}
John Toner and Yuhai Tu.
\newblock {Flocks, herds, and schools: A quantitative theory of flocking}.
\newblock {\em Physical Review E}, 58(4):4828--4858, 1998.

\bibitem{Couzin2003}
Iain~D Couzin and Jens Krause.
\newblock {Self-Organization and Collective Behavior in Vertebrates}.
\newblock {\em Advances in the study of behavior}, 32:1--75, 2003.

\bibitem{Couzin2002}
Iain~D Couzin, Jens Krause, Richard James, Graeme~D Ruxton, and Nigel~R Franks.
\newblock {Collective Memory and Spatial Sorting in Animal Groups}.
\newblock pages 1--11, 2002.

\bibitem{Guillaume2004}
Gregoire Guillaume and Hugues Chate.
\newblock {Onset of Collective and Cohesive Motion}.
\newblock {\em Physical Review Letters}, 92(2), 2004.

\bibitem{Zafeiris}
Tamas Vicsek and Anna Zafeiris.
\newblock {Collective motion}.
\newblock {\em Physics Reports}, 517(3-4):71--140, 2012.

\bibitem{Chepizhko2013b}
Oleksandr Chepizhko, Eduardo~G Altmann, and Fernando Peruani.
\newblock {Optimal noise maximizes collective motion in heterogeneous media}.
\newblock {\em arXiv}, page 1305.5707v1, 2013.

\bibitem{McCandlish2012}
Samuel~R McCandlish, Aparna Baskaran, and Michael~F Hagan.
\newblock {Spontaneous segregation of self-propelled particles with different
  motilities}.
\newblock {\em arXiv}, page 1110.2479v2, 2012.

\bibitem{Baglietto2013}
Gabriel Baglietto, EV~Albano, and J~Candia.
\newblock {Gregarious vs individualistic behavior in Vicsek swarms and the
  onset of first-order phase transitions}.
\newblock {\em Physica A: Statistical Mechanics and its Applications}, pages
  1--13, 2013.

\bibitem{DOrsogna2006}
M.~D$\backslash$’Orsogna, Y.~Chuang, A.~Bertozzi, and L.~Chayes.
\newblock {Self-Propelled Particles with Soft-Core Interactions: Patterns,
  Stability, and Collapse}.
\newblock {\em Physical Review Letters}, 96(10):104302, mar 2006.

\bibitem{Gazi2004}
Veysel Gazi and Kevin~M. Passino.
\newblock {A class of attractions/repulsion functions for stable swarm
  aggregations}.
\newblock {\em International Journal of Control}, 77(18):1567--1579, dec 2004.

\bibitem{Couzin2005}
Iain~D Couzin, Jens Krause, Nigel~R Franks, and Simon~a Levin.
\newblock {Effective leadership and decision-making in animal groups on the
  move.}
\newblock {\em Nature}, 433(7025):513--6, feb 2005.

\bibitem{Belmonte2008}
Julio Belmonte, Gilberto Thomas, Leonardo Brunnet, Rita de~Almeida, and Hugues
  Chat{\'{e}}.
\newblock {Self-Propelled Particle Model for Cell-Sorting Phenomena}.
\newblock {\em Physical Review Letters}, 100(24):248702, jun 2008.

\bibitem{Quint}
David~A Quint and Ajay Gopinathan.
\newblock {Topologically induced swarming phase transition on a 2D percolated
  lattice}.
\newblock {\em Physical Biology}, 0(0):0, 2015.

\bibitem{Toner2005}
John Toner, Yuhai Tu, and Sriram Ramaswamy.
\newblock {Hydrodynamics and phases of flocks}.
\newblock {\em Annals of Physics}, 318(1):170--244, jul 2005.

\bibitem{Toner1995}
J~Toner and Y~Tu.
\newblock {Long-Range Order in a Two-Dimensional Dynamical XY Model: How Birds
  Fly Together}.
\newblock {\em Physical Review Letters}, 75(23):4326--4329, 1995.

\bibitem{Marchetti2013}
M.~C. Marchetti, J.~F. Joanny, S.~Ramaswamy, T.~B. Liverpool, J.~Prost, Rao
  Madan, and R.~{Aditi Simha}.
\newblock {Hydrodynamics of soft active matter}.
\newblock {\em Reviews of Modern Physics}, 85, 2013.

\bibitem{Levine2000}
Herbert Levine, Wouter-Jan Rappel, and Inon Cohen.
\newblock {Self-organization in systems of self-propelled particles}.
\newblock {\em Physical Review E}, 63(1):017101, dec 2000.

\bibitem{Bertin2009}
Eric Bertin, Michel Droz, and Guillaume Gr{\'{e}}goire.
\newblock {Hydrodynamic equations for self-propelled particles: microscopic
  derivation and stability analysis}.
\newblock {\em Journal of Physics A: Mathematical and Theoretical}, 42:445001,
  2009.

\bibitem{Liu2013}
Quan-Xing Liu, Arjen Doelman, Vivi Rottsch{\"{a}}fer, Monique de~Jager, Peter
  M~J Herman, Max Rietkerk, and Johan van~de Koppel.
\newblock {Phase separation explains a new class of self-organized spatial
  patterns in ecological systems.}
\newblock {\em Proceedings of the National Academy of Sciences of the United
  States of America}, 110(29):11905--10, jul 2013.

\bibitem{Cavagna2010}
A.~Cavagna, A.~Cimarelli, I.~Giardina, G.~Parisi, R.~Santagati, F.~Stefanini,
  and M.~Viale.
\newblock {Scale-free correlations in starling flocks}.
\newblock {\em Proceedings of the National Academy of Sciences},
  107(26):11865--11870, 2010.

\bibitem{Gautrais2009}
Jacques Gautrais, Christian Jost, Marc Soria, Alexandre Campo, S{\'{e}}bastien
  Motsch, Richard Fournier, St{\'{e}}phane Blanco, and Guy Theraulaz.
\newblock {Analyzing fish movement as a persistent turning walker}.
\newblock {\em Journal of Mathematical Biology}, 58(3):429--445, 2009.

\bibitem{Doxzen2013}
Kevin Doxzen, Sri Ram~Krishna Vedula, Man~Chun Leong, Hiroaki Hirata, Nir~S.
  Gov, Alexandre~J. Kabla, Benoit Ladoux, and Chwee~Teck Lim.
\newblock {Guidance of collective cell migration by substrate geometry}.
\newblock {\em Integrative Biology}, 5(8):1026, 2013.

\end{thebibliography}
\end{document}